\documentclass[12pts]{article}

\usepackage{amsmath,amssymb,amsthm,epsf,epsfig}
\usepackage{amsmath,amssymb}
\usepackage{rotating}
\begin{document}

\begin{center}
\huge {\bf A generalized quantum nonlinear oscillator}
\end{center}
\vspace{.7 cm}

\begin{center}
{ B.Midya \footnote {E-mail: bikash.midya@gmail.com}, B.
Roy\footnote {E-mail: barnana@isical.ac.in}
~ \\ Physics \& Applied Mathematics Unit\\ Indian Statistical
Institute \\ Kolkata  700108\\ India}
\end{center}

\vspace{1 cm}

\begin{center} {\large {{\bf Abstract}}} \end{center}
We examine various generalizations e.g, exactly solvable, quasi exactly solvable and
non Hermitian variants of the quantum nonlinear oscillator. For all these cases the same mass
function has been used and it has also been shown that the new exactly solvable potentials
posses shape invariance symmetry. The solutions are obtained in terms of
classical orthogonal polynomials.

\vspace{1 cm}

\section {\bf Introduction}
Recently there has been a surge of interest in obtaining exact
{\cite{r1}} and quasi exact solution {\cite{r2}}
 of the position dependent mass Schr\"{o}dinger equation (PDMSE) for various potentials and mass functions by
 using various methods like Lie algebraic techniques {\cite{r3}}, supersymmetric quantum mechanics (factorization method)
  {\cite{r4,r5}}, shape invariance approach {\cite{r6}}, point canonical transformation {\cite{r7}}, path integral formalism {\cite{r8}},
  transfer matrix method {\cite{r9}} etc. Apart from the intrinsic interest the motivation behind this issue arises because of the relevance of position dependent mass in describing the physics of many microstructures of current interest, such as compositionally graded crystals {\cite{r30}},
 quantum dots {\cite{r31}}, $^3$He clusters {\cite{r32}}, metal clusters {\cite{r33}} etc. The concept of position
  dependent mass comes from the effective mass approximation \cite{r34} which is an useful tool for studying the
  motion of carrier electrons in pure crystals and also for the virtual-crystal approximation in the treatment of
  homogeneous alloys (where the actual potential is approximated by a periodic potential) as well as in graded mixed
  semiconductors (where the potential is not periodic). The attention to the effective mass approach stems from the
  extraordinary development in crystallographic growth techniques which allow the production of non uniform semiconductor
  specimen with abrupt heterojunctions. In these mesoscopic materials, the effective mass of the charge carriers are
  position dependent.  Consequently the study of the effective mass Schr\"{o}dinger equation becomes relevant for deeper
   understanding of the non-trivial quantum effects observed on these nanostructures. The position dependent (effective)
   mass is also used in the construction of pseudo-potentials which have a significant computational advantage
   in quantum Monte Carlo method \cite{r35}. It has also been found that such equations appear in very different areas. For example, it has been shown that constant
  mass Schr\"odinger equation in curved space and those based on deformed commutation relations can be interpreted in
  terms of PDMSE in flat space \cite{r14} and $\cal {PT}$-symmetric cubic anharmonic oscillator \cite{r15}.

The nonlinear differential
equation
\begin{equation}
(1+\lambda x^2)\ddot {x} - (\lambda x)\dot {x}^2 + \alpha^2 x =
0,~~~\lambda > 0 \label{eqn1}
\end{equation}
was studied by Mathews and Lakshmanan in \cite{r16,r29} as an
example of a non-linear oscillator and it was shown that the
solution of (\ref{eqn1}) is
\begin{equation}
x = A sin(\omega t + \phi)\label{sol1}
\end{equation}
with the following additional restriction linking frequency and
amplitude
\begin{equation}
\omega^2 = \frac{\alpha^2}{1+\lambda A^2}\label{free}
\end{equation}
%That is, the equation(1) represents a non-linear oscillator with periodic solutions that were qualified as having a simple harmonic form.
Furthermore (\ref{eqn1}) can be obtained from the Lagrangian
\cite{r16}
\begin{equation}
{L} = \frac{1}{2}\frac{1}{(1+\lambda x^2)}(\dot {x}^2 - \alpha^2
x^2)\label{lag}
\end{equation}
so that both the kinetic and the potential term depend on the same
parameter $\lambda$. So this nonlinear oscillator must be
considered as a particular case of a system with a position
dependent effective mass. Recently in a series of papers
\cite{r17,r18} this particular nonlinear system has been
generalized to the higher dimensions and various properties of
this system have been studied. The classical Hamiltonian
corresponding to the $\lambda$-dependent oscillator is given by
\cite{r16,r18}

\begin{equation}
H = \left(\frac{1}{2m}\right)P_x^2 +
\left(\frac{1}{2}\right)g\left(\frac{x^2}{1+\lambda
x^2}\right),~~~~~ P_x = \sqrt{1+\lambda
x^2}p_x,~~g=m\alpha^2\label{e45}
\end{equation}
$p_x$ being the canonically conjugate momentum defined by $p_x =
\frac{\partial L}{\partial \dot{x}}$, ${L}$ being the Lagrangian and $m$ is the mass.

It has been shown in \cite{r18} that in the space ${\cal{L}}^2(\Re
,d\mu)$ where $d\mu = (\frac{1}{\sqrt{1+\lambda x^2}})dx$, the
differential operator $\sqrt{1+\lambda x^2}\frac{d}{dx}$ is skew
self adjoint. Therefore,
 contrary to the naive expectation of ordering ambiguities, the transition from the classical system to
 the quantum one is given by defining the momentum operator
\begin{equation}
P_x = -i \sqrt{1+\lambda x^2}\frac{d}{dx}
\end{equation}
so that

$$
(1+\lambda x^2)p_x^2 \rightarrow -\left(\sqrt{1+\lambda x^2}\frac{d}{dx}\right)
\left(\sqrt{1+\lambda x^2}\frac{d}{dx}\right)$$

Therefore the quantum version of the Hamiltonian (\ref{e45}) with
$\hbar=1$ becomes \cite{r18}
\begin{equation}
\hat {H} = -\frac{1}{2m}(1+\lambda x^2)\frac{d^2}{dx^2} - \left(\frac{1}{2m}\right)\lambda x\frac{d}{dx}
 + \frac{1}{2}g\left(\frac{1}{1+\lambda x^2}\right)\label{hamil1}
\end{equation}
where $g = \alpha (m\alpha + \lambda)$. It is to be
 noted that in ref \cite{r18} the value of the parameter $g$ has been slightly modified from that given in equation (\ref{e45}).

It may be pointed out that
this $\lambda$-dependent system can be considered as a deformation of the standard harmonic
oscillator in the sense that for $\lambda \rightarrow 0$ all the characteristics of the linear oscillator are recovered.

In ref \cite{r18}, the PDMSE corresponding to this nonlinear
oscillator has been solved exactly as a Sturm-Liouville
 problem and $\lambda$-dependent eigenvalues and eigenfunctions were
obtained for both $\lambda > 0 $ and $\lambda < 0$. The
$\lambda$-dependent wave functions were shown to be related to a
family of orthogonal polynomials
 that can be considered as $\lambda$-deformations of the standard Hermite polynomials.
 Also the Schr\"odinger factorization formalism, intertwining method and shape invariance
 approach were discussed with reference to this particular quantum Hamiltonian. The existence of a
$\lambda$-dependent Rodrigues formula, a generating function and $\lambda$-dependent recursion relations were obtained.\\
In this paper our objective is to re-examine this problem and
obtain closed form expression for the normalisation constant,
modified generating function and recursion relations for $\Lambda
(= \frac{\lambda}{\alpha})$-deformed Hermite polynomials. A
relation between the $\Lambda$ deformed Hermite polynomials and
Jacobi polynomials will also be obtained. We shall also obtain a
number of exactly solvable, quasi exactly solvable and non
Hermitian potentials corresponding to the same mass function
$m(x)=(1+\lambda x^2)^{-1}$. It will be seen that some of these
potentials are generalizations of the nonlinear oscillator
potentials while the others are of different types. It will
be shown that these exactly solvable potentials are shape invariant.
Moreover these potentials can also be complexified and by doing so we shall
also obtain a number of exactly solvable non Hermitian potentials
within the framework of PDMSE. As a method of obtaining these
results we shall use point canonical transformation consisting of
change of coordinate only. The organization of the paper is as follows: in section
\ref{sol} we shall obtain exactly solvable potentials and a
relation between $\Lambda$-deformed Hermite polynomials and Jacobi
polynomials; in section \ref{shape} it is shown that the exactly
solvable potentials are shape invariant; in section \ref{complex}
we obtain exactly solvable non Hermitian potentials; section
\ref{cqes} deals with complex quasi exactly solvable potentials
and finally section \ref{dis} is devoted to a discussion.\\

\section {\bf Exactly solvable potentials for the mass $m(x) = \left(\frac{1}{1+\lambda
x^2}\right)$}\label{sol}

 Here we shall obtain exact solutions
PDMSE for a number of potentials with the same
 mass function $m(x)=\left(\frac{1}{1+\lambda x^2}\right)$.
 For this purpose we first write the PDMSE corresponding to the Hamiltonian given in Eqn.(\ref{hamil1}) with $m=1$ and $\lambda > 0$ as
\begin{equation}
\left[-(1+\lambda
    x^2)\frac{d^2\psi}{dx^2}-\lambda x
    \frac{d\psi}{dx} - \frac{g}{\lambda}\left(\frac{1}{1+\lambda x^2}\right)\right]\psi = E
    \psi\label{e5}
\end{equation}
\begin{equation}
E = 2e - \frac{g}{\lambda}\label{e43}
\end{equation}
where $e$ is the energy for the Hamiltonian (\ref{hamil1}). Now
expanding $(1+\lambda x^2)^{-1}$ for
$|x|<\frac{1}{\sqrt{\lambda}}$ we can write the potential of
equation (\ref{e5}) as
 \begin{equation}V(x)=-\frac{g}{\lambda}+g
x^2-\lambda~ O(x^3)\label{e44}
\end{equation} It is clear from (\ref{e44}) that the term
$(-\frac{g}{\lambda})$ in equation (\ref{e43}) cancels from both
sides of the equation (\ref{e5}), so that the new eigenvalues
(\ref{e43}) are actually the old eigenvalues $e$ of the
Hamiltonian (\ref{hamil1}). Also, as $\lambda\rightarrow 0$, the
potential and the eigenvalues of equation (\ref{e5}) reduces to
those of a linear harmonic oscillator.

 Now generalizing the
potential of the  equation (\ref{e5}) as below, the corresponding
PDMSE now reads
    \begin{equation}
    -(1+\lambda
    x^2)\frac{d^2\psi}{dx^2}-\lambda x
    \frac{d\psi}{dx}+\left[\frac{B^2-A^2-A\sqrt{\lambda}}{1+\lambda x^2}+B(2A+\sqrt{\lambda}) \left(\frac{\sqrt{\lambda}x}
    {1+\lambda x^2}\right)+A^2\right]\psi=E\psi\label{eigen2}
    \end{equation}
It is seen from (\ref{eigen2}) that if we put $B=0$ then the
potential reduces to that of the nonlinear oscillator with
$\frac{g}{\lambda} = A^2 + A\sqrt{\lambda}$. It is to be
noted that this generalization should correctly reproduce the $\lambda\rightarrow 0$ limit,
in which case the equation (\ref{eigen2})
reduces to the Schr\"{o}dinger equation for linear harmonic oscillator.
In Appendix $\bf{I}$  we have shown that in the limit
$\lambda\rightarrow 0$ and for $A=\frac{\alpha}{\sqrt{\lambda}}$
(which is one of the solution of the quadratic equation
$A^2+A\sqrt{\lambda}=\frac{g}{\lambda}$), $B=0$ the potential of
equation (\ref{eigen2}), the energy eigenvalues (\ref{en2}) and
the wavefunction given in ({\ref{e8}}) reduces to those of a
linear harmonic oscillator. This particular generalization is made
so that it corresponds to the hyperbolic Scarf II potential
{\cite{r19}} in the constant mass case. In order to solve
(\ref{eigen2}), we now perform a transformation involving
change of variable given
by
\begin{equation} z = \int \frac{dx}{\sqrt{F(x)}} =
\frac{1}{\sqrt{\lambda}}~sinh^{-1}(\sqrt{\lambda}x)\label{t}
\end{equation}
where
\begin{equation}
\begin{array}{lcl}
F(x)&=&1+\lambda x^2~~,~~\lambda > 0\\
\end{array}\label{e6}
\end{equation}
Under the transformation (\ref{t}), Eqn.(\ref{eigen2}) reduces to a Schr\"odinger equation
\begin{equation}
-\frac{d^2\psi}{dz^2} + V(z)\psi(z) = E\psi(z)\label{rsch}
\end{equation}
where the potential $V(z)$ is given by
\begin{equation}
V(z) =
(B^2-A^2-A\sqrt{\lambda})sech^{2}\left(z\sqrt{\lambda}\right) +
B(2A+\sqrt{\lambda})~tanh\left(z\sqrt{\lambda}\right)sech\left(z\sqrt{\lambda}\right)+A^2~~~\label{pot1}
\end{equation}
The potential (\ref{pot1}) is a standard solvable potential and
the solutions are given by {\cite{r19}}

    \begin{equation}
     \psi_{n}(z)=N_n {i^n} \left(1+sinh
     ^{2}(z\sqrt{\lambda})\right)^{-\frac{s}{2}} e^{-r
     tan^{-1}\left(sinh(z\sqrt{\lambda})\right)}~P_{n}^{(ir-s-\frac{1}{2},-ir-s-\frac{1}{2})}
     \left(i~sinh(z\sqrt{\lambda}\right)\label{e7}
     \end{equation}
 where $N_n$ is the normalization constant ,
  $s=\frac{A}{\sqrt{\lambda}},~ ~r=\frac{B}{\sqrt{\lambda}}$ and $P_n^{(\alpha,\beta)}(x)$ is
   the Jacobi Polynomial {\cite{r21}}.
The normalization constants $N_n, n=0,1,2,...$ are given by
{\cite{r20}},
    \begin{equation}N_n=\left[\frac{ \sqrt{\lambda}~ n!~(s-n)\Gamma(s-ir-n+\frac{1}{2})\Gamma{(s+ir-n+\frac{1}{2})}}
    {\pi~2^{-2s}\Gamma(2s-n+1)}\right]^{1/2}\label{e40}
    \end{equation}
The eigenvalues $E_n$ are given by
    \begin{equation}
    E_n=n\sqrt{\lambda}(2A-n\sqrt{\lambda}),~~~~~
    n=0,1,2,...<s\label{en2}
    \end{equation}
 Subsequently by performing the inverse of the transformation (\ref{t}) we find the
 solution of PDMSE (\ref{eigen2}) as
    \begin{equation}
    \begin{array}{ll}
    \displaystyle \psi_{n}(x)=\left[\frac{\sqrt{\lambda} ~n!~(s-n)\Gamma(s-ir-n+\frac{1}{2})\Gamma{(s+ir-n+\frac{1}{2})}}
    {\pi~2^{-2s}\Gamma(2s-n+1)}\right]^{1/2}\\
    \displaystyle ~~~~~~~~~~~
    i^n (1+\lambda x^2)^{-\frac{s}{2}} e^{-r tan^{-1}(x\sqrt{\lambda})}
    P_n^{(ir-s-\frac{1}{2},-ir-s-\frac{1}{2})}(ix\sqrt{\lambda})~~~,~~~~~~ n=0,1,2,\cdots <s~(=\frac{A}{\sqrt{\lambda}})
    \end{array}\label{e8}
    \end{equation}
At this point it is natural to ask the following question :
 Are there other solvable potentials corresponding to the mass function $m(x)=\left(\frac{1}{1+\lambda x^2}\right)$?
  The answer to this question is in the affirmative. The procedure to obtain these potentials is similar
   and so instead of treating each case separately we have presented the potentials and the corresponding solutions
 in Table 1. The first two and the last two potentials in Table 1. are actually the generalizations of the nonlinear oscillator potential. Although the other two potentials in the Table are not generalizations of the nonlinear oscillator potential, nevertheless they are exactly solvable potentials with the same mass
 function.\\

\subsection{\bf Relation between $\Lambda$-deformed Hermite polynomial and Jacobi polynomial, Generating function, Recursion relation}
Here we shall obtain a correspondence between the
$\Lambda$-deformed Hermite polynomials {\cite{r18}} and Jacobi
polynomials. We recall that the Hamiltonian for nonlinear
oscillator is given by {\cite{r18}}
$$
H = - \frac{1}{2}(1+ \lambda x^2)\frac{d^2}{dx^2} -
\frac{1}{2}\lambda x \frac{d}{dx} + \frac{g}{2}\left(\frac{x^2}{1+
\lambda x^2}\right)
$$
After introducing adimensional variables $(y,\Lambda)$ as was done
in ref {\cite{r18}}
\begin{equation}
y = \sqrt{\alpha} x ~~,~~ \Lambda =
\frac{\lambda}{\alpha}\label{e42}
\end{equation}
the Schr\"odinger equation $H\psi = \epsilon \psi$ reduces to
\begin{equation}
\left[- \frac{1}{2}(1+ \Lambda y^2)\frac{d^2}{dy^2} -
\frac{1}{2}\Lambda y \frac{d}{dy} +
\frac{1+\Lambda}{2}\left(\frac{y^2}{1+ \Lambda
y^2}\right)\right]\psi = \epsilon \psi \label{eqn}
\end{equation}
The eigenvalues and eigenfunctions for $\Lambda < 0$ are
\cite{r18}
\begin{equation}
\begin{array}{lcl}
\psi_m(y,\Lambda) &=& {\cal {H}}_m(y,\Lambda)(1 - |\Lambda|y^2)^{\frac{1}{(2|\Lambda|)}}\\
\epsilon_m &=& (m+\frac{1}{2}) - \frac{1}{2}m^2 \Lambda~~~,~~~m =
0,1,2,...
\end{array} \label{eqn5}
\end{equation}
where ${\cal {H}}_m(y,\Lambda)$ is $\Lambda$-deformed Hermite
polynomial whose Rodrigues formula and generating function are
given in (\ref{rf}). For $\Lambda
> 0$,
\begin{equation}
\begin{array}{lcl}
\psi_m(y,\Lambda) &=& {\cal {H}}_m(y,\Lambda)(1+\Lambda y^2)^{-\frac{1}{2\Lambda}}\\
\epsilon_m &=& (m+\frac{1}{2}) - \frac{1}{2}m^2\Lambda~~~,~~~m = 0,1,2
\cdots,N_{\Lambda}
\end{array}\label{eqn6}
\end{equation}
where $N_{\Lambda}$ denotes the greatest integer lower than
$m_{\Lambda}(=\frac{1}{\Lambda})$. On the other hand, putting $B =
0$ and $A = \frac{\alpha}{\sqrt{\lambda}}$ in the solution
(\ref{e8}) of Eqn.(\ref{eigen2}), the eigenfunctions of
Eqn.(\ref{eqn}) can be written in terms of Jacobi polynomial as
\begin{equation}
\psi_n(y) = N_n(1+\Lambda
y^2)^{-\frac{1}{2\Lambda}}P_n^{(-\frac{1}{2}-\frac{1}{\Lambda},-\frac{1}{2}-\frac{1}{\Lambda})}(iy\sqrt{\Lambda})~~,~~n
= 0,1,2\cdots <\frac{1}{\Lambda}~~~(\Lambda > 0) \label{eqn7}
\end{equation}
For $\Lambda < 0$, putting $B=0,
A=\frac{\alpha}{\sqrt{|\lambda|}}$ in the wavefunction of the 5th
entry of  Table 1. and using (\ref{e42}) we obtain
\begin{equation}
\psi_n(y) = N_n(1+\Lambda y^2)^{-\frac{1}{2\Lambda}}
P_n^{(-\frac{1}{2}-\frac{1}{\Lambda},-\frac{1}{2}-\frac{1}{\Lambda})}(y\sqrt{|\Lambda|})~~,
~~n = 0,1,2\cdots ~~~(\Lambda < 0) \label{eqn8}
\end{equation}
Comparing Eqns.(\ref{eqn5}) and (\ref{eqn8}) and also
Eqns.(\ref{eqn6}) and (\ref{eqn7}), it is possible to derive a
relation between $\Lambda$-deformed Hermite polynomial ${\cal
{H}}_n(y,\Lambda)$ and Jacobi polynomial $P_n^{(\alpha,
\beta)}(x)$ as
\begin{equation}
P_n^{(-\frac{1}{2}-\frac{1}{\Lambda},-\frac{1}{2}-\frac{1}{\Lambda})}(iy\sqrt{\Lambda})
= \frac{1}{n!}\left(\frac{1}{2i\sqrt{\Lambda}}\right)^n {\cal
{H}}_n(y,\Lambda),~~~~~~\forall~ \Lambda \label{e9}
\end{equation}
The Rodrigues formula and the generating function for the
$\Lambda$-deformed Hermite polynomial ${\cal {H}}_n(y,\Lambda)$
were given by {\cite{r18}}
\begin{equation}
\begin{array}{lcl}
{\cal {H}}_n(y,\Lambda) &=& (-1)^nz_y^{\frac{1}{\Lambda}+\frac{1}{2}}\frac{d^n}{dy^n}\left[z_y^n z_y^{-(\frac{1}{\Lambda}+\frac{1}{2})}\right],~~~z_y = 1 + \Lambda y^2\\
{\cal {F}}(t,y,\Lambda) &=& (1+\Lambda (2ty -
t^2))^{\frac{1}{\Lambda}}
\end{array}\label{rf}
\end{equation}
It was shown {\cite{r18}} that the polynomials obtained from the
generating function ${\cal {F}}(t,y,\Lambda)$ with those obtained
from Rodrigues formula are essentially the same and only differ in
the values of the global multiplicative coefficients. We have
observed that if the generating function ${\cal {F}}(t,y,\Lambda)$
is taken as
 \begin{equation}
    (1+\Lambda(2ty-t^2))^{\frac{1}{\Lambda}}=\sum_{n=0}^\infty
    \frac{1}{2^n}
      \frac{\left(-\frac{1}{\Lambda}\right)_n}{\left(\frac{1}{2}-\frac{1}{\Lambda}\right)_n}
      {\cal {H}}_n(y,\Lambda)\frac{t^n}{n!}\label{e10}
    \end{equation}
where $(a)_n$ represents P\"ochhammer symbol given by
$(a)_n=\frac{\Gamma(a+n)}{\Gamma(a)}$ then the polynomials
obtained from the above relation are exactly
same with those obtained from Rodrigues formula given in Eqn.(\ref{rf}).\\
Correspondingly the recursion relations are obtained as
\begin{equation}
    (\Lambda(2n+1)-2)~ [2(1-n\Lambda) y {\cal {H}}_n(y,\Lambda)+(\Lambda(2n-1)-2)n
    {\cal {H}}_{n-1}(y,\Lambda)]=(n\Lambda-2){\cal
    {H}}_{n+1}(y,\Lambda)\label{rr1}
\end{equation}
and
\begin{equation}
\begin{array}{ll}
\displaystyle (\Lambda(n-2)-2) ~[2(\Lambda(2n-1)-2)n
    {\cal {H}}_n(y,\Lambda)-(\Lambda(n-1)-2){\cal {H}}'_n(y,\Lambda)]\\
    \displaystyle
    ~~~~~=n\Lambda(\Lambda(2n-1)-2)~[2(\Lambda(n-2)-2)y{\cal {H}}'_{n-1}(y,\Lambda)-(n-1)
    (\Lambda(2n-3)-2){\cal {H}}'_{n-2}(y,\Lambda)]
    \end{array}\label{rr2}
    \end{equation}
    where `prime' denotes differentiation with respect to $y$.
    For $\Lambda \rightarrow 0$ Eqns.(\ref{rr1}) and (\ref{rr2}) give the recursion relations for Hermite
    polynomial {\cite{r21}}.\\

\section{\bf Shape invariance approach to supersymmetric PDMSE}\label{shape}
Supersymmetric approach to PDMSE {\cite{r5}} may be discussed
either by reducing the PDMSE to constant mass Schr\"odinger
equation or start with modified intertwining operators consisting
of first order differential operators. Here we shall be following
the later approach. Thus we consider operators of the form
\begin{equation}
A = P_x - i W(x),~~~~   A^\dagger = P_x + i W(x),~~~~ P_x =
\frac{1}{\sqrt{m(x)}}\left(-i\frac{d}{dx}\right) \label{op}
\end{equation}
We now consider the supercharges $Q, Q^\dagger$ defined by
\begin{equation}
Q=\begin{pmatrix}
      0 & 0 \\
      A & 0 \\
    \end{pmatrix}~, ~~~~
    Q^\dagger  =  \begin{pmatrix}
      0 & A^\dagger \\
      0 & 0 \\
    \end{pmatrix}\label{e11}
    \end{equation}
The supersymmetric Hamiltonian is then obtained as
\begin{equation}
H^{PDM} = \{Q,Q^{\dagger}\} = \begin{pmatrix}
                                   H_-^{PDM} & 0 \\
                                   0   & H_+^{PDM} \\
                                   \end{pmatrix}
                                  = \begin{pmatrix} A^\dagger A & 0 \\
                                   0   & AA^\dagger \\
                                   \end{pmatrix}\label{susy3}
\end{equation}
where the component Hamiltonians are given by
\begin{equation}
H_{\pm}^{PDM} = -\frac{1}{m(x)}\frac{d^2}{dx^2} +
\left(\frac{m^\prime}{2m^2}\right)\frac{d}{dx} + W^2 \pm
\frac{W^\prime}{\sqrt{m}}\label{e12}
\end{equation}
The Hamiltonians $H_{\pm}^{PDM}$ are supersymmetric partners and
the potentials are
\begin{equation}
V_{\pm}^{PDM} = W^2(x) \pm \frac{W'(x)}{\sqrt{m(x)}}\label{e1}
\end{equation}
It can be easily seen that the following commutation and
anticommutation relations
\begin{equation}
\begin{array}{lcl}
Q^2 = {Q^{\dagger}}^2 = [Q,H^{PDM}] = [Q^\dagger,H^{PDM}] &=& 0\\
\{Q,Q^{\dagger}\} = \{Q^{\dagger},Q\} &=& 0
\end{array}\label{e13}
\end{equation}
together with Eqn.(\ref{susy3}) complete the standard
supersymmetry algebra {\cite{r19,r22}}. For unbroken Supersymmetry (SUSY), the
ground state of $H_-$ has zero energy $(E_0^{(-)} = 0)$ provided
the ground state wave function $\psi_0^{(-)}(z)$ given by
($A\psi_0^{(-)} = 0$)
\begin{equation}
\psi_0^{(-)}(x) = N_0 ~exp~\left[-\int^{x}~\sqrt{m(y)}~
W(y)dy\right] \label{gs}
\end{equation}
is normalizable. In this case it can be shown that, apart from the ground
state of $H_-$, the partner Hamiltonians $H_{\pm}$ have identical
bound-state spectra. In particular, they satisfy
\begin{equation}
E_{n+1}^{(-)} = E_n^{(+)}~~,n=0,1,2 \cdots \label{e14}
\end{equation}
The eigenfunctions of $H_{\pm}$ corresponding to the same
eigenvalue are related by
\begin{equation}
\begin{array}{lcl}
A\psi_{n+1}^{(-)} &=& \left(E_n^{(+)}\right)^{\frac{1}{2}}\psi_n^{(+)}(x)\\
A^{\dagger}\psi_n^{(+)}(x) &=&
\left(E_n^{(+)}\right)^{\frac{1}{2}}\psi_{n+1}^{(-)}(x)
\end{array}\label{e15}
\end{equation}
It may be noted here that the superpotential $W(x)$ and therefore the factorization of the Hamiltonian
could be generated from the ground state solution of the  equation.
%This in turn led to the fact that there are a hierarchy of related Hamiltonians, $H_1$,$H_2$,$H_3$,$\cdots$ $H_n$, having the same bound state spectra except $m$ states are missing and $0<m<N$, where $N+1$ is the number of bound states of $H_1$.
In a remarkable paper {\cite{r23}}, Gendenshtein explored the
relationship between SUSY, and solvable potentials.
 The pair of potentials $V_{\pm}(x,a_0)$, $a_0$ being a set of parameters, are called
 shape invariant if they satisfy the relationship {\cite{r5,r19}}
\begin{equation}
\begin{array}{lcl}
V_+(x,a_0) &=& W^2(x,a_0) + W'(x,a_0)\\
         &=& W^2(x,a_1) - W'(x,a_1) + R(a_0)\\
         &=& V\_(x,a_1) + R(a_0)
\end{array}\label{e2}
\end{equation}
where $a_1$ is some function of $a_0$ and $R(a_0)$ is independent
of $x$. When SUSY is unbroken the energy spectrum of any shape
invariant potential is given by {\cite{r19}}
\begin{equation}
E^{(-)}_n=\sum_{i=0}^{n-1} R(a_i)~;~~E^{(-)}_0=0\label{e16}
\end{equation}
 We are now going to study
the factorization and the shape invariance property of the
 potentials for the PDMSE Schr\"{o}dinger equation.
As an example let us consider the generalized nonlinear oscillator
of section \ref{sol}. For this it is now necessary to choose the
superpotential $W(x)$ so that $H_-$ can be identified with the
Hamiltonian of Eqn.(\ref{eigen2}). In this case we choose the
superpotential to be
\begin{equation}
W = A \frac{\sqrt{\lambda}x}{\sqrt{1+\lambda x^2}} +
B\frac{1}{\sqrt{1+\lambda x^2}}\label{e17}
\end{equation}
Therefore the Hamiltonians $H_-^{PDM}$ and $H_+^{PDM}$ can be
factorised as
\begin{equation}
\begin{array}{lcl}
H_-^{PDM} &=& A^{\dagger}A\\
          &=& -(1+\lambda x^2)\frac{d^2}{dx^2} - \lambda x \frac{d}{dx} + \frac{B^2-A^2-A\sqrt{\lambda}}{1+\lambda x^2}+
           B(2A + \sqrt{\lambda})(\frac{\sqrt{\lambda}x}{1+\lambda x^2}) + A^2\\
H_+^{PDM} &=& AA^{\dagger}\\
          &=& -(1+\lambda x^2)\frac{d^2}{dx^2} - \lambda x \frac{d}{dx} + \frac{B^2-A^2+A\sqrt{\lambda}}{1+\lambda x^2}+ B(2A - \sqrt{\lambda})(\frac{\sqrt{\lambda}x}{1+\lambda x^2}) + A^2
\end{array}\label{e18}
\end{equation}
These two Hamiltonians are related by
\begin{equation}
H_+^{PDM}(x;A,B) = H_-^{PDM}(x;A-\sqrt{\lambda},B) +
\sqrt{\lambda}(2A-\sqrt{\lambda})\label{e19}
\end{equation}
so that they satisfy shape invariance condition
\begin{equation}
H_+^{PDM}(x,a_0) = H_-^{PDM}(x,a_1) + R(a_0)\label{e20}
\end{equation}
where $\{a_0\} = (A,B), \{a_1\} = (A-\sqrt{\lambda},B)$ and
$R(a_0) = \sqrt{\lambda}(2A-\sqrt{\lambda}).$

The ground state $\psi_0(x,a_0)$ of the Hamiltonian $H_-^{PDM}$ is
found by solving $A\psi_0(x,a_0)=0$, and has a zero energy i.e.
\begin{equation}
H_-^{PDM}(x,a_0)\psi_0(x,a_0)=0\label{e47}
 \end{equation} Now using (\ref{e20}) we
can see that $\psi_0(x,a_1)$ is an eigenstate of $H_+^{PDM}$ with
the energy $E_1=R(a_0)$, because
\begin{equation}\begin{array}{ll}\displaystyle
H_+^{PDM}(x,a_0)\psi_0(x,a_1)=H_-^{PDM}(x,a_1)\psi_0(x,a_1)+R(a_0)\psi_0(x,a_1)\\
\displaystyle
~~~~~~~~~~~~~~~~~~~~~~~~~~~=R(a_0)\psi_0(x,a_1)~,~~~\mbox{[using
({\ref{e47}})]}\label{e46}
\end{array}
\end{equation}
Next, using the intertwining relation
$H_-^{PDM}(x,a_0)A^{\dagger}(x,a_0)=A^{\dagger}(x,a_0)H_+^{PDM}(x,a_0)$
and equation ({\ref{e20}}), we see that
\begin{equation}
H_-^{PDM}(x,a_0)A^{\dagger}(x,a_0)\psi_0(x,a_1)=A^{\dagger}(x,a_0)H_+^{PDM}(x,a_0)\psi_0(x,a_1)=A^{\dagger}
\left[H_-^{PDM}(x,a_1)+R(a_0)\right]\psi_0(x,a_1)\label{e48}
\end{equation}
and hence using (\ref{e47}) we arrive at
\begin{equation}
H_-^{PDM}(x,a_0)A^{\dagger}(x,a_0)\psi_0(x,a_1)=
R(a_1)A^{\dagger}(x,a_0)\psi_0(x,a_1)\label{e49}
\end{equation}
This indicates that $A^{\dagger}(x,a_0)\psi_0(x,a_1)$ is an
eigenstate of $H_-^{PDM}$ with an energy $E_1=R(a_0)$.
 Now iterating this
process we will find the sequence of energies for $H_-^{PDM}$ as
\begin{equation}
E^{(-)}_n = \sum_{i=0}^{n-1} R(a_i) =
n\sqrt{\lambda}(2A-n\sqrt{\lambda})~, ~~~E^{(-)}_0=0\label{e21}
\end{equation}
and corresponding eigenfunctions being
\begin{equation}
\psi_n(x,a_0)=A^{\dagger}(x,a_0)A^{\dagger}(x,a_1)...A^{\dagger}(x,a_{n-1})\psi_0(x,a_n)\label{e50}
\end{equation}
where $a_i=f(a_{i-1})=\underbrace{f(f(....(f(a_0)))}_{i \mbox{
times}}=\left(A-i\sqrt{\lambda},B\right)$ and
$R(a_i)=\sqrt{\lambda}\left[2\left(A-i\sqrt{\lambda}\right)-\sqrt{\lambda}\right]$.

We have found a number of other potentials which are shape
invariant for the same mass function. For all these potentials the
energy, wave functions and other parameters related to shape
invariance property are given in Table 1.\\

\subsection{Shape invariance approach to PDMSE with broken supersymmetry}

When supersymmetry is broken neither of the wave functions
$\psi_0^{(\pm)}(x)\approx exp[\pm\int^x\sqrt{m(y)}W(y)dy]$ are
normalizable and in this case all the energy values are degenerate
i.e, $H_+$ and $H_-$ have identical energy eigenvalues
{\cite{r19,r23}}
\begin{equation}
E_n^{(-)} = E_n^{(+)}\label{e22}
\end{equation}
with ground state energies greater than zero. So far as we know,
little attention has been paid till now to study problems
involving broken SUSY in the case of PDMSE. Broken supersymmetric
shape invariant systems in the case of constant mass Schr\"odinger
equation has been discussed in ref {\cite{r24}}. Below we
illustrate the two step procedure discussed in {\cite{r25}} for
obtaining the energy spectra in PDMSE when the SUSY is broken. \
For this, we consider the superpotential as
    \begin{equation}
    W(x,A,B)=A\sqrt{|\lambda|}\frac{x}{\sqrt{1+\lambda
    x^2}}-\frac{B}{\sqrt{|\lambda|}}\frac{\sqrt{1+\lambda x^2}}{x}, ~~~~~ 0<x
    <\frac{1}{\sqrt{|\lambda|}},~~\lambda<0. \label{sp}
    \end{equation}
 Then the supersymmetric partner potentials are obtained using (\ref{e1}) as
    \begin{equation}
    \begin{array}{ll}
    \displaystyle
    V_-(x,A,B)=\frac{A(A-\sqrt{|\lambda|})}{1+\lambda
    x^2}-\frac{B(B-\sqrt{|\lambda|})}{\lambda x^2}-(A+B)^2   \\
    \displaystyle
    V_+(x,A,B)=\frac{A(A+\sqrt{|\lambda|})}{1+\lambda
    x^2}-\frac{B(B+\sqrt{|\lambda|})}{\lambda x^2}-(A+B)^2
    \end{array}\label{bsp}
    \end{equation}
The ground state wave function is obtained from (\ref{gs}) as
    \begin{equation}
    \psi_0^{(-)}\sim x^{\frac{B}{\sqrt{|\lambda|}}}~ (1+\lambda
    x^2)^{\frac{A}{2\sqrt{|\lambda|}}}\label{e23}
    \end{equation}
For $A>0,~B>0$ the ground state wave function $\psi_0^{(-)}$ is
normalizable which means the SUSY is unbroken. But for $A>0,~
B<0$  and $A<0,~ B>0$ , neither of $\psi_0^{(\pm)}$ are normalizable. Hence SUSY  is broken in both cases. \\
We shall discuss the case $A>0,~ B<0$. In this case the eigenstates of $V_\pm (x,A,B)$ are related by
    \begin{equation}
    \begin{array}{lll}
    \displaystyle
    \psi_n^{(+)}(x,a_0)=A(x,a_0) \psi_n^{(-)} (x,a_0)\\
    \displaystyle \psi_n^{(-)}(x,a_0)=A^\dagger (x,a_0)
    \psi_n^{(+)}(x,a_0),\\
    \displaystyle E_n^{(-)}(a_0)=E_n^{(+)}(a_0)
    \end{array}\label{bse1}
    \end{equation}
 Now we can show the potentials in equation (\ref{bsp}) are shape invariant
 by two different relations between the parameters.\\

{\bf {Step 1}}\\
 The potentials of equation (\ref{bsp}) are shape invariant if we
 change $A\rightarrow A+\sqrt{|\lambda|}~ \mbox{and} ~ B \rightarrow
 B+\sqrt{|\lambda|}.$The shape invariant condition is given by
    \begin{equation}
    V_+(x,A,B)=V_-~\left(x,A+\sqrt{|\lambda|},B+\sqrt{|\lambda|}\right)+\left(A+B+2\sqrt{|\lambda|}\right)^2-(A+B)^2\label{e24}
    \end{equation}
 Now for $B<-\frac{1}{\sqrt{|\lambda|}}$ it is seen that the superpotential
  (\ref{sp}) resulting from change of parameters as above falls in the class of broken SUSY
problem for which $E_0^{(-)} \neq 0$. Though the potentials of
equation (\ref{bsp}) are shape invariant but we are unable to
determine the spectra for these potentials because of the absence
of zero energy ground state.\\
Another way of parameterizations $A\rightarrow A+\sqrt{|\lambda|}~
~\mbox{and}~~ B\rightarrow -B$ gives us
    \begin{equation}
        V_+(x,A,B)=V_-~\left(x,A+\sqrt{|\lambda|},-B\right)+\left(A-B+\sqrt{|\lambda|}\right)^2-(A+B)^2
        \label{sic1}
    \end{equation}
which shows that $V_-~\mbox{and}~ V_+$ are shape invariant. This
change of parameters $(A\rightarrow A+\sqrt{|\lambda|}~
~\mbox{and}~~ B\rightarrow -B)$ leads to a system with unbroken
SUSY since the parameter $B$ changes sign.
 Hence the ground state energy of the potential
$V_-(x,A+\sqrt{|\lambda|},-B)$ is zero. From the relation
(\ref{sic1}) we observe that $V_+(x,A,B) ~\mbox{and} ~
V_-\left(A+\sqrt{|\lambda|},-B\right)$ differ only by a constant,
hence we have
    \begin{equation}
    \begin{array}{ll}
    \displaystyle
    \psi_+(x,A,B)=\psi_-(x,A+\sqrt{|\lambda|},-B)\\
    \displaystyle
    E_n^{(+)}(A,B)=E_n^{(-)}(x,A+\sqrt{|\lambda|},-B)+\left(A-B+\sqrt{|\lambda|}\right)^2-(A+B)^2
    \end{array}\label{bse2}
    \end{equation}
Thus, if we can evaluate the spectrum and energy eigenfunctions of
unbroken SUSY $H_-^{PDM} (x,A+\sqrt{|\lambda|},-B)$, then we can
determine the spectrum and eigenfunctions $H_+^{PDM}(x,A,B)$ with broken SUSY. In the 2nd step we will do this.\\

{\bf{Step 2}}\\
With the help of shape invariant formalism in case of unbroken SUSY
for PDMSE  (See section [3]) we obtain spectrum and eigenfunctions
for $V_-(x,A+\sqrt{|\lambda|},-B)$ as
    \begin{equation}\begin{array}{ll}\displaystyle
    E_n^{(-)}(A+\sqrt{|\lambda|},-B)=\left(A-B+\sqrt{|\lambda|}+2n\sqrt{|\lambda|}\right)^2-
    \left(A-B+\sqrt{|\lambda|}\right)^2\\
    \displaystyle
    \psi_n^{(-)}(x,A+\sqrt{|\lambda|},-B) \propto x^{\frac{B}{\sqrt{|\lambda|}}}(1+\lambda
    x^2)^{\frac{A}{2\sqrt{|\lambda|}}} P_n^{(\frac{B}{\sqrt{|\lambda|}}-\frac{1}{2},\frac{A}{\sqrt{|\lambda|}}-\frac{1}{2})}
    (1+2\lambda x^2)
    \end{array}\label{bse3}
    \end{equation}
Now using (\ref{bse3}) ,(\ref{bse2}) and (\ref{bse1}) we obtain
spectrum and eigenfunctions for $V^- (x,A,B)$ with broken SUSY as
    \begin{equation}
    \begin{array}{ll}
    \displaystyle
        E_n^{(-)}(A,B)=\left(A-B+\sqrt{|\lambda|}+2n\sqrt{|\lambda|}\right)^2-(A+B)^2\\
        \displaystyle
        \psi_n^{(-)}(x,A,B) \propto x^{\frac{1-B}{\sqrt{|\lambda|}}}(1+\lambda x^2)^{\frac{A}{2\sqrt{|\lambda|}}}
         P_n^{\left(\frac{1}{2}-\frac{B}{\sqrt{|\lambda|}},\frac{A}{\sqrt{|\lambda|}}-\frac{1}{2}\right)}(1+2\lambda x^2)
         \end{array}\label{e25}
         \end{equation}
Similar approach can be applied in case of $A<0~ \mbox{and}~ B>0$.
In this case we change $(A,B)$ into $(-A,B+\sqrt{|\lambda|})$ and
the shape invariance condition is
\begin{equation}
V_+(x,A,B)=V_-~\left(x,-A,B+\sqrt{|\lambda|}\right)+\left(B-A+\sqrt{|\lambda|}\right)^2-(A+B)^2\label{e26}
\end{equation}
And
\begin{equation}
\begin{array}{ll}
\displaystyle
E_n^{(-)}(A,B)=\left(B-A+\sqrt{|\lambda|}+2n\sqrt{|\lambda|}\right)^2-(A+B)^2\\
\displaystyle
\psi_n^{(-)}(x,A,B) \propto x^{\frac{1-A}{\sqrt{|\lambda|}}}(1+\lambda
x^2)^{\frac{B}{2\sqrt{|\lambda|}}}
P_n^{\left(\frac{B}{\sqrt{|\lambda|}}-\frac{1}{2},\frac{A}{\sqrt{|\lambda|}}-\frac{1}{2}\right)}(1+2\lambda
x^2)
\end{array}\label{e27}
\end{equation}

\section {\bf Exactly solvable $\cal{PT}$ symmetric potentials in PDMSE}\label{complex}
Here we shall find exactly solvable complex potentials, some of
which are related to the nonlinear oscillator potential, within
the framework of PDMSE. Before we consider any particular
potential let us note that a quantum mechanical Hamiltonian $H$ is
said to be $\cal {PT}$ symmetric {\cite{r25}} if
\begin{equation}
\cal {PT} H = H \cal {PT}\label{e28}
\end{equation}
where $\cal {P}$ is the parity operator acting as spatial
reflection, and $\cal {T}$ stands for time reversal, acting as the
complex conjugation operator. Their action on the position and
momentum operators are given by
\begin{equation}
{\cal {P}}: x \rightarrow -x,~~p \rightarrow -p,~~~~{\cal {T}}: x \rightarrow x, p \rightarrow -p,
i \rightarrow -i\label{pt}
\end{equation}
For a constant mass Schr\"odinger Hamiltonian the condition for $\cal{PT}$ symmetry reduces to $V(x)=V^*(-x)$.
However in the case of position dependent mass an additional condition is required. To see this we note that in the
present case the Hamiltonian is of the form
\begin{equation}
H = -\frac{1}{2m(x)}\frac{d^2}{dx^2}-\frac{m^\prime(x)}{2m^2(x)}\frac{d}{dx}+V(x)\label{hamil3}
\end{equation}
From (\ref{pt}) it follows that the conditions for the Hamiltonian (\ref{hamil3}) to be $\cal{PT}$ symmetric are
\begin{equation}
m(x) = m(-x)~~,~~V(x) = V^*(-x)\label{mv}
\end{equation}
It may be pointed out that here we are working with a mass profile
$m(x)=(1+\lambda x^2)^{-1}$ which is an even function and
consequently satisfies the first condition of (\ref{mv}). To
generate non-Hermitian interaction in the present case we introduce complex coupling constant. As an example let us first consider the
potential appearing in (\ref{eigen2}). It can be seen from
(\ref{en2})
 that the energy for this potential does not depend on one of the potential parameters, namely $B$.
  Thus we consider the complex potential
\begin{equation}
V(x) = \left[\frac{B^2-A^2-A\sqrt{\lambda}}{1+\lambda x^2}+iB(2A+\sqrt{\lambda}) \left(\frac{\sqrt{\lambda}x}
    {1+\lambda x^2}\right)+A^2\right]\label{ptv}
\end{equation}
From (\ref{ptv}) it can be easily verified that $V(x)=V^*(-x)$ so
that the Hamiltonian (\ref{hamil3}) with this potential is
$\cal{PT}$ symmetric. In this case the spectrum is real and given
by (\ref{en2}). Proceeding in a similar way we have obtained the
spectrum of a number of $\cal{PT}$ symmetric potentials and the
results are given in Table 2. Incidentally all the potentials in
Table 2 are shape invariant and the results can also be obtained
algebraically.\\

\section{\bf Quasi exactly solvable \cal {PT} symmetric potentials in PDMSE}\label{cqes}
The complex sextic potential in the constant mass Schr\"{o}dinger
equation has been discussed in ref {\cite{r27}}. By using the
transformations (\ref{t}) for $\lambda > 0$ we obtain the corresponding quasi exactly solvable potentials in PDMSE.\\
For $\lambda > 0$, the potential is taken as
\begin{equation}
V(x)=\sum_{k=1}^{6}\frac{c_k}{\lambda^{\frac{k}{2}}}{\left(sinh(x\sqrt{\lambda})\right)^{-k}}\label{qesp}
\end{equation}
where for $V(x)$ to be $\cal{PT}$ symmetric, $c_1, c_3, c_5$ are purely imaginary and $c_2, c_4, c_6$ are real. \\
Following ref {\cite{r27}}, the ansatz for the wave function is
taken as
\begin{equation}
\psi(x)=f(x)~ exp\left(-\sum_{j=1}^{4}
\frac{b_j}{\lambda^{\frac{j}{2}}}
(sinh(x\sqrt{\lambda}))^{-j}\right) \label{qesw}
\end{equation}
where $f(x)$ is some polynomial function of $x$.
We shall focus on the following choices of $f(x)$:
$$ \hspace{-7 cm}(a)~~~ f(x) = 1$$
$$ \hspace{-4 cm}(b)~~~ f(x) = \frac{\left(sinh(x\sqrt{\lambda})\right)^{-1}}{\sqrt{\lambda}} + a_0$$
$$(c)~~~f(x)=\frac{\left(sinh(x\sqrt{\lambda})\right)^{-2}}{\lambda}
+a_1\frac{\left(sinh(x\sqrt{\lambda})\right)^{-1}}{\sqrt{\lambda}}+a_0$$
For complex potentials, $a_0$ is purely imaginary in (b), but in (c) $a_1$ is purely imaginary, but $a_0$ is real.\\
Without going into the details of calculation, which are quite straightforward, let us summarize our results.\\

{\bf Case 1:} $f(x) = 1$\\
In this case the relation between the parameters $c_i$ and $b_i$
are found to be
\begin{equation}
c_1 = -3b_3+2b_1b_2,~~c_2 = -6b_4+3b_1b_3+2b_2^2,~~c_3 = 4b_1b_4+6b_2b_3\\
c_4 = 8b_2b_4+\frac{9}{2}b_3^2,~~ c_5 = 12b_3b_4,~~c_6 =
8b_4^2\label{e38}
\end{equation}
and
\begin{equation}
E=b_2-\frac{1}{2}b_1^2\label{e37}
\end{equation}
Without loss of generality, we can choose $c_6 = \frac{1}{2}$
which fixes the leading coefficient of $V(x)$. It gives $b_4 = \pm
\frac{1}{4}$. Taking the positive sign to ensure the
normalizability of the wave function we obtain
\begin{equation}
\psi(x)=exp~\left(-\frac{b_{1}\left(sinh(x\sqrt{\lambda})\right)^{-1}}{\sqrt{\lambda}}
-\frac{b_{2}\left(sinh(x\sqrt{\lambda})\right)^{-2}}{\lambda}-\frac{b_{3}\left(sinh(x\sqrt{\lambda})\right)^{-3}}
{\lambda\sqrt{\lambda}}
-\frac{\left(sinh(x\sqrt{\lambda})\right)^{-4}}{4\lambda^2}\right)\label{e36}
\end{equation}
Now if $b_1$ and $b_3$ are purely imaginary then $c_1, c_3, c_5$ are also purely imaginary.
In that case $V(x)$ in Eqn.(\ref{qesp}) and $\psi(x)$ in Eqn.(\ref{qesw}) are $\cal {PT}$ symmetric and $E$ is real.\\

{\bf{Case 2:}} $f(x)=\displaystyle \frac{\left(sinh(x\sqrt{\lambda})\right)^{-1}}{\sqrt{\lambda}} + a_0$, where $a_0$ is
purely imaginary.\\
In this case wave function is of the form
\begin{equation}
\begin{array}{ll} \displaystyle
\psi(x)=\left(\frac{\left(sinh(x\sqrt{\lambda})\right)^{-1}}{\sqrt{\lambda}}
+a_0\right)\\
\displaystyle
~~~~~~~~~exp\left(-\frac{b_{1}\left(sinh(x\sqrt{\lambda})\right)^{-1}}{\sqrt{\lambda}}
-\frac{b_{2}\left(sinh(x\sqrt{\lambda})\right)^{-2}}{\lambda}-\frac{b_{3}\left(sinh(x\sqrt{\lambda})\right)^{-3}}
{\lambda\sqrt{\lambda}}
-\frac{\left(sinh(x\sqrt{\lambda})\right)^{-1}}{4\lambda^2}\right)
\end{array}\label{e35}
\end{equation}
In this case the relation between the parameters are given by
\begin{equation}
c_1=-6b_3+2b_1 b_2+a_0~~,~~c_2=-\frac{5}{2}+3b_1b_3+2b_2^2~~,~~c_3=b_1+6b_2b_3\\
c_4=2b_2+\frac{9}{2}b_3^2~~,~~c_5=3b_3~~,~~c_6=\frac{1}{2}\label{e34}
\end{equation}
$a_0$ satisfies the condition
\begin{equation}
a_0^3-3b_3a_0^2+2b_2a_0-b_1=0\label{qesc}
\end{equation}
The energy is given by
\begin{equation}
E=-\frac{1}{2}b_1^2+3b_2-3a_0 b_3+a_0^2\label{e33}
\end{equation}
We now consider two special cases.\\
(a) $b_1=b_3=0$ and $a_0^2<0$.\\
    In this case $c_1$ is purely imaginary and $c_3 = c_5 = 0$. Moreover $c_1=a_0=\pm i\sqrt{2b_2}$.
So we get two different complex potentials corresponding to above
two values of $c_1$ with same real energy eigenvalues. The
potential, energyvalues and the eigenfunctions are given by
\begin{equation}
\begin{array}{lcl}
V(x) &=& \frac{1}{2}\frac{(sinh(x\sqrt{\lambda}))^{-6}}{\lambda^3}+\frac{2b_2}
{\lambda^2}(sinh(x\sqrt{\lambda}))^{-4}+\frac{(2b_2^2-\frac{5}{2})}{\lambda}(sinh(x\sqrt{\lambda}))^{-2}\pm
\frac{i \sqrt{2b_2}}{\sqrt{\lambda}}\left(sinh(x\sqrt{\lambda})\right)^{-1}\\
E &=& b_2>0\\
\psi(x) &=&
\left(\frac{(sinh(x\sqrt{\lambda})^{-1}}{\sqrt{\lambda}}\pm i
\sqrt{2b_2}\right)
exp\left(-\frac{b_2}{\lambda}\left(sinh(x\sqrt{\lambda})\right)^{-2}
-\frac{1}{4\lambda ^2}
\left(sinh(x\sqrt{\lambda})\right)^{-4}\right)
\end{array}\label{e32}
\end{equation}
It can be easily seen from the above equations that the potential
is $\cal {PT}$ symmetric,
 while the wave function is odd under $\cal {PT}$ symmetry.\\

(b) $b_1=0, b_3\neq 0$\\
Then from (\ref{qesc}) we get
\begin{equation}
a_0=\frac{1}{2} (3b_3 \pm \sqrt{9b_3^2-8b_2})\label{e31}
\end{equation}
So in order to make $a_0$ imaginary we must
have $9b_3^2-8b_2<0.$ or $b_3^2 = -|b_3|^2 \leq \frac{8}{9}b_2$.\\
In this case also there exist two different complex potentials
corresponding to two values of $b_3$ with the same real energy
eigenvalues $E=3b_2-3a_0b_3+a_0^2.$

 {\bf{Case 3:}}
$f(x)=\frac{\left(sinh(x\sqrt{\lambda})\right)^{-2}}{\lambda} +a_1
\frac{\left(sinh(x\sqrt{\lambda})\right)^{-1}}{\sqrt{\lambda}}+a_0$,
where $a_1$ is imaginary and $a_0$ is real.\\
In this case the relation between the parameters is given by
 \begin{equation}
a_1 = 2b_3~~,~~ a_0 = \frac{1}{2} \left(2b_2 - b_3^2 \pm
\sqrt{(2b_2-3b_3^2)^2 + 2} \right)\label{qesc1}
 \end{equation}
The wave function, energy and the potential are of the form
\begin{equation}
    \begin{array}{ll}
    \psi_\pm(x)=\left[\frac{\left(sinh(x\sqrt{\lambda})\right)^{-2}}{\lambda}+2b_3 \frac{\left(sinh(x\sqrt{\lambda})\right)^{-1}}{\sqrt{\lambda}}+\frac{1}{2}
    (2b_2-b_3^2\pm \sqrt{(2b_2-3b_3^2)^2+2}) \right]\\
    \displaystyle
      ~~~~~~exp \left(-2b_3(b_2-b_3^2)\frac{\left(sinh(x\sqrt{\lambda})\right)^{-1}}{\sqrt{\lambda}}
    -b_2 \frac{\left(sinh(x\sqrt{\lambda})\right)^{-2}}{\lambda}
    -b_3\frac{\left(sinh(x\sqrt{\lambda})\right)^{-3}}{\lambda\sqrt{\lambda}}
    -\frac{1}{4}\frac{\left(sinh(x\sqrt{\lambda})\right)^{-4}}{\lambda^2}\right)
    \end{array}\label{e30}
    \end{equation}
    \begin{equation}
    E_\pm=-2b_3^2(b_2-b_3^2)^2+3b_2-b_3^2\pm
    \sqrt{(2b_2-3b_3^2)^2+2}\label{e29}
    \end{equation}
    \begin{equation}
    \begin{array}{ll}
    \displaystyle
    V(x) =\frac{1}{2\lambda^3}(sinh(x\sqrt{\lambda}))^{-6}+\frac{3b_3}{\lambda^2 \sqrt{\lambda}}(sinh(x\sqrt{\lambda}))^{-5}+\frac{(2b_2+\frac{9}{2})}{\lambda ^2}(sinh(x\sqrt{\lambda}))^{-4}+\frac{2b_3}{\lambda \sqrt{\lambda}}(4b_2-b_3^2)(sinh(x\sqrt{\lambda}))^{-3}\\
\displaystyle
~~~~~~~~~~~+\frac{[2(b_2^2+3b_2b_3^2-3b_3^4)-\frac{7}{2}]}{\lambda}(sinh(x\sqrt{\lambda}))^{-2}
+\frac{b_3(4b_2^2-4b_2b_3^2-7)}{\sqrt{\lambda}}(sinh(x\sqrt{\lambda}))^{-1}
\end{array}\label{qesv2}
    \end{equation}
The result (\ref{qesc1}) to (\ref{qesv2}) are valid both for real
and purely imaginary $b_i$. When $b_i$ are purely imaginary the
potential and wave function are $\cal {PT}$ symmetric while for
real $b_i$ $\cal {PT}$- symmetry is broken. In particular when
$b_3$ is purely imaginary we have a complex $\cal {PT}$- symmetric
two parameter family of potentials corresponding to two values of
$a_0$ with two distinct
real eigenvalues.\\

\section{\bf Discussion}\label{dis}
We have studied various exactly solvable as well as quasi exactly solvable and
non-Hermitian generalizations of the quantum nonlinear oscillator with the mass function
$\left(\frac{1}{1+\lambda x^2}\right)$. We have also obtained a closed form normalisation
constant for the eigenfunctions of quantum nonlinear oscillator. A relationship between the $\lambda$
deformed Hermite polynomial and Jacobi polynomial has also been found. By exploiting supersymmetry of the PDMSE
we have obtained some shape invariant potentials corresponding to this particular mass function. We have considered the
shape invariance approach to PDMSE with broken supersymmetry as well. As for the future work we feel it would be
interesting to examine Lie algebraic symmetry of the exactly solvable potentials. In view of the fact that in the present
 case the transformation (\ref{t}) is invertible, it seems promising to study whether or not the Lie algebraic symmetry
 of the constant mass system can be transported back to the non constant mass case. Another interesting area of
 investigation would be to study the classical analogs of some of the models (especially the $\cal{PT}$ symmetric ones)
  considered here.\\
  \newpage
{\begin{sidewaystable}
    \centering
    {\tiny{
    \begin{tabular}{|l c| c |c |c |c |c |}
    \hline

    & $V(x)$ & $W(x)$& $E_n$ & $\psi_n(x)$ & $a_{i},i=0,1,..$ & $R(a_i)$\\

    \hline
    \hline

    &$\frac{B^2-A^2-A\sqrt{\lambda}}{{1+\lambda x^2}}+B(2A+\sqrt{\lambda})\frac{\sqrt{\lambda}x}{1+\lambda
    x^2}+A^2$~~~
    &$A\frac{\sqrt{\lambda}x}{\sqrt{1+\lambda x^2}}+B\frac{1}{\sqrt{1+\lambda
    x^2}}$~~~& $n\sqrt{\lambda}(2A-n\sqrt{\lambda})$ ~~~& $i^n (1+\lambda x^2)^{-\frac{s}{2}}e^{-r~tan^{-1}
    (x\sqrt{\lambda})}$& $(A-i\sqrt{\lambda},B)$&$\sqrt{\lambda}\left[2A-(2i+1)\sqrt{\lambda}\right]$\\
    &$$&$$&$$ &$P_n^{(ir-s-\frac{1}{2},-ir-s-\frac{1}{2})}(ix\sqrt{\lambda})$&$$&$$\\

    \hline

    &$A^2+\frac{B^2}{A^2}-\frac{A(A+\sqrt{\lambda})}{1+\lambda x^2}+2B\frac{\sqrt{\lambda}x}{\sqrt{1+\lambda x^2}}~,~ B<A^2$
    &$A\frac{\sqrt{\lambda }x}{\sqrt{1+\lambda x^2}}+\frac{B}{A}$~~~& $A^2+\frac{B^2}{A^2}-
    (A-n\sqrt{\lambda})^2-\frac{B^2}{(A-n\sqrt{\lambda})^2}$ ~~~
    &$\left(1-\frac{x\sqrt{\lambda}}{\sqrt{1+\lambda
    x^2}}\right)^{\frac{s_1}{2}}\left(1+\frac{x\sqrt{\lambda}}{\sqrt{1+\lambda
    x^2}}\right)^{\frac{s_2}{2}}
    $&$(A-i\sqrt{\lambda},B)$&$A^2-\left[A-(i+1)\sqrt{\lambda}\right]^2$\\&$$&$$ &$$ & $P_n^{(s_1,s_2)}\left(\frac{x\sqrt{\lambda}}{\sqrt{1+\lambda
    x^2}}\right)$&$$&$+\frac{B^2}{A^2}-\frac{B^2}{\left[A-(i+1)\sqrt{\lambda}\right]^2}$\\

    \hline

    &$A^2+\frac{B^2}{A^2}-2B\frac{\sqrt{1+\lambda x^2}}{\sqrt{\lambda}x}+\frac{A(A-\sqrt{\lambda})}{\lambda
    x^2}~,~B>A^2$
    &$\frac{B}{A}-A\frac{\sqrt{1+\lambda x^2}}{x\sqrt{\lambda}}$
    & $A^2+\frac{B^2}{A^2}-
    (A+n\sqrt{\lambda})^2-\frac{B^2}{(A+n\sqrt{\lambda})^2}$& $\left(\frac{\sqrt{1+\lambda x^2}}{x\sqrt{\lambda}}
    -1\right)^
    {\frac{s_3}{2}}\left(\frac{\sqrt{1+\lambda x^2}}{x\sqrt{\lambda}}+1\right)^{\frac{s_4}{2}}$
    &$(A+i\sqrt{\lambda},B)$&$A^2-\left[A+(i+1)\sqrt{\lambda}\right]^2$\\
    &$0\leq x\sqrt{\lambda}\leq \infty$& $$ &$$ &$P_n^{(s_3,s_4)}\left(\frac{1+\lambda x^2}{x\sqrt{\lambda}}\right)$&$$
    &$+\frac{B^2}{A^2}-\frac{B^2}{\left[A+(i+1)\sqrt{\lambda}\right]^2}$\\

    \hline

    & $\frac{A^2+B^2+A\sqrt{\lambda}}{\lambda x^2}-B(2A+\lambda)\frac{\sqrt{1+\lambda x^2}}{\lambda
    x^2}+A^2~,~A<B$&$A\frac{\sqrt{1+\lambda
    x^2}}{x\sqrt{\lambda}}-B\frac{1}{x\sqrt{\lambda}}$&$n\sqrt{\lambda}(2A-n\sqrt{\lambda})$&$(\sqrt{1+\lambda
    x^2}-1)^{(\frac{r-s}{2})}
    (\sqrt{1+\lambda x^2}+1)^{-(\frac{r+s}{2})}$&$(A-i\sqrt{\lambda},B)$&$\sqrt{\lambda}\left[2A-(2i+1)\sqrt{\lambda}\right]$\\
    &$0\leq x\sqrt{\lambda}\leq \infty$&$$&$$&$P_n^{(r-s-\frac{1}{2},-r-s-\frac{1}{2})}\left(\sqrt{1+\lambda
    x^2}\right)$&$$&$$\\

    \hline

    &$\frac{A^2+B^2-A\sqrt{|\lambda|}}{1+\lambda x^2}-B(2A-\sqrt{|\lambda|})
    \frac{x\sqrt{|\lambda|}}{1+\lambda x^2}-A^2$&$A\frac{x\sqrt{|\lambda|}}{\sqrt{1+\lambda x^2}}
    -B\frac{1}{\sqrt{1+\lambda
    x^2}}$&$n\sqrt{|\lambda|}(2A+n\sqrt{|\lambda|})$&$(1-x\sqrt{|\lambda|})^{(\frac{s'-r'}{2})}
    (1+x\sqrt{|\lambda|})^{(\frac{r'+s'}{2})}$&$(A+i\sqrt{|\lambda|},B)$&$\sqrt{|\lambda|}
    \left[2A+(2i+1)\sqrt{|\lambda|}\right]$\\
    &$\frac{1}{-\sqrt{|\lambda|}}\leq
    x\leq\frac{1}{\sqrt{|\lambda|}}$&$$&$$&$P_n^{(s'-r'-\frac{1}{2},s'+r'-\frac{1}{2})}\left(x\sqrt{|\lambda|}\right)$&$$&$$\\

    \hline

    &$\frac{A(A-\sqrt{|\lambda|})}{1+\lambda x^2}-2B\frac{x\sqrt{|\lambda|}}{1+\lambda
    x^2}-A^2+\frac{B^2}{A^2}$&$A\frac{x\sqrt{|\lambda|}}{\sqrt{1+\lambda
    x^2}}-\frac{B}{A}$&$\frac{B^2}{A^2}-A^2+(A+n\sqrt{|\lambda|})^2-\frac{B^2}{(A+n\sqrt{|\lambda|})^2}$
    &$\left(\frac{\lambda x^2}{1+\lambda
    x^2}-1\right)^{-(\frac{s'+n}{2})}e^{-a\sqrt{|\lambda|}x}$&$(A+i\sqrt{|\lambda|},B)$&$\frac{B^2}{A^2}-\frac{B^2}{\left[A+(i+1)\sqrt{|\lambda|}\right]^2}$\\
    &$\frac{1}{-\sqrt{|\lambda|}}\leq
    x\leq\frac{1}{\sqrt{|\lambda|}}$&$$&$$
    &$P_n^{(-s'-n-ia,-s'-n+ia)}\left(-i\frac{x\sqrt{|\lambda|}}{\sqrt{1+\lambda x^2}}\right)$&$$&$-A^2+\left[A+(i+1)\sqrt{|\lambda|}\right]^2$\\

    \hline

      \end{tabular}
      }}
       \caption { \small{Exactly solvable shape invariant potentials $V(x)$, superpotential $W(x)$ , energy eigenvalue
      $E_n$ and wave functions $\psi_n(x).$ Where $s=\frac{A}{\sqrt{\lambda}}, r=\frac{B}{\sqrt{\lambda}},
       r_1=\frac{B}{\lambda}, a=\frac{r_1}{s-n}, s_1=s-n+a,s_2=s-n-a, s_3=a-n-s,s_4=-(s+n+a), s'=\frac{A}
       {\sqrt{|\lambda|}} ~\&~r'=\frac{B}{\sqrt{|\lambda|}}.$The first four entries correspond to $\lambda>0$ and the last two
        correspond to $\lambda<0.$}}

    \end{sidewaystable}}
{\begin{sidewaystable}
    \centering
    {\tiny{
    \begin{tabular}{|l c| c |c |c|}
    \hline

    & $V(x)$ & $W(x)$& $E_n$ & $\psi_n(x)$ \\

    \hline
    \hline

    &$\frac{-B^2-A^2-A\sqrt{\lambda}}{1+\lambda x^2}+iB(2A+\sqrt{\lambda})\frac{\sqrt{\lambda}x}{1+\lambda
    x^2}+A^2$~~~
    &$A\frac{\sqrt{\lambda}x}{\sqrt{1+\lambda x^2}}+iB\frac{1}{\sqrt{1+\lambda
    x^2}}$~~~& $n\sqrt{\lambda}(2A-n\sqrt{\lambda})$ ~~~& $i^n (1+\lambda x^2)^{-\frac{s}{2}}e^{-r~tan^{-1}
    (x\sqrt{\lambda})}$\\&$$&$$&$$ &$P_n^{(ir-s-\frac{1}{2},-ir-s-\frac{1}{2})}\left(ix\sqrt{\lambda}\right)$\\

    \hline

    &$A^2-\frac{B^2}{A^2}-\frac{A(A+\sqrt{\lambda})}{1+\lambda x^2}+i2B\frac{\sqrt{\lambda}x}{\sqrt{1+\lambda x^2}}~,~ B<A^2$
    &$A\frac{\sqrt{\lambda }x}{\sqrt{1+\lambda x^2}}+i\frac{B}{A}$~~~& $A^2-\frac{B^2}{A^2}-
    (A-n\sqrt{\lambda})^2+\frac{B^2}{(A-n\sqrt{\lambda})^2}$ ~~~
    &$\left(1-\frac{x\sqrt{\lambda}}{\sqrt{1+\lambda
    x^2}}\right)^{\frac{s_1}{2}}\left(1+\frac{x\sqrt{\lambda}}{\sqrt{1+\lambda
    x^2}}\right)^{\frac{s_2}{2}}
    $\\&$$&$$ &$$ & $P_n^{(s_1,s_2)}\left(\frac{x\sqrt{\lambda}}{\sqrt{1+\lambda
    x^2}}\right)$\\

    \hline

    &$A^2-\frac{B^2}{A^2}-2iB\frac{\sqrt{1+\lambda x^2}}{\sqrt{\lambda}x}+\frac{A(A-\sqrt{\lambda})}{\lambda
    x^2}~,~B>A^2$
    &$i\frac{B}{A}-A\frac{\sqrt{1+\lambda x^2}}{x\sqrt{\lambda}}$
    & $A^2-\frac{B^2}{A^2}-
    (A+n\sqrt{\lambda})^2+\frac{B^2}{(A+n\sqrt{\lambda})^2}$& $\left(\frac{\sqrt{1+\lambda x^2}}{x\sqrt{\lambda}}
    -1\right)^
    {\frac{s_3}{2}}\left(\frac{\sqrt{1+\lambda x^2}}{x\sqrt{\lambda}}+1\right)^{\frac{s_4}{2}}$\\
    &$0\leq x\sqrt{\lambda}\leq \infty$& $$ &$$ &$P_n^{(s_3,s_4)}\left(\frac{1+\lambda x^2}{x\sqrt{\lambda}}\right)$\\

    \hline

    & $\frac{A^2-B^2+A\sqrt{\lambda}}{\lambda x^2}-iB(2A+\lambda)\frac{\sqrt{1+\lambda x^2}}{\lambda
    x^2}A^2~,~A<B$&$A\frac{\sqrt{1+\lambda
    x^2}}{x\sqrt{\lambda}}-iB\frac{1}{x\sqrt{\lambda}}$&$n\sqrt{\lambda}(2A-n\sqrt{\lambda})$&$(\sqrt{1+\lambda
    x^2}-1)^{(\frac{r-s}{2})}
    (\sqrt{1+\lambda x^2}+1)^{-(\frac{r+s}{2})}$\\
    &$0\leq x\sqrt{\lambda}\leq \infty$&$$&$$&$P_n^{(r-s-\frac{1}{2},-r-s-\frac{1}{2})}\left(\sqrt{1+\lambda
    x^2}\right)$\\

    \hline

    &$\frac{A^2-B^2-A\sqrt{|\lambda|}}{1+\lambda x^2}-iB(2A-\sqrt{|\lambda|})
    \frac{x\sqrt{|\lambda|}}{1+\lambda x^2}-A^2$&$A\frac{x\sqrt{|\lambda|}}{\sqrt{1+\lambda x^2}}
    -iB\frac{1}{\sqrt{1+\lambda
    x^2}}$&$n\sqrt{\lambda}(2A+n\sqrt{|\lambda|})$&$(1-x\sqrt{|\lambda|})^{(\frac{s'-r'}{2})}
    (1+x\sqrt{|\lambda|})^{(\frac{r'+s'}{2})}$\\
    &$\frac{1}{-\sqrt{|\lambda|}}\leq
    x\leq\frac{1}{\sqrt{|\lambda|}}$&$$&$$&$P_n^{(s'-r'-\frac{1}{2},s'+r'-\frac{1}{2})}\left(x\sqrt{|\lambda|}\right)$\\

    \hline

    &$\frac{A(A-\sqrt{|\lambda|})}{1+\lambda x^2}-2iB\frac{x\sqrt{|\lambda|}}{1+\lambda
    x^2}-A^2-\frac{B^2}{A^2}$&$A\frac{x\sqrt{|\lambda|}}{\sqrt{1+\lambda
    x^2}}-i\frac{B}{A}$&$-\frac{B^2}{A^2}-A^2+(A+n\sqrt{|\lambda|})^2+\frac{B^2}{(A+n\sqrt{|\lambda|})^2}$
    &$\left(\frac{\lambda x^2}{1+\lambda
    x^2}-1\right)^{-(\frac{s'+n}{2})}e^{-a\sqrt{|\lambda|}x}$\\
    &$\frac{1}{-\sqrt{|\lambda|}}\leq
    x\leq\frac{1}{\sqrt{|\lambda|}}$&$$&$$
    &$P_n^{(-s'-n-ia,-s'-n+ia)}\left(-i\frac{x\sqrt{|\lambda|}}{\sqrt{1+\lambda x^2}}\right)$\\

    \hline

      \end{tabular}
      }}
      \caption {\small{ Exactly solvable PT Symmetric potentials, where $s=\frac{A}{\sqrt{\lambda}}, r=i\frac{B}{\sqrt{\lambda}},
       r_1=i\frac{B}{\lambda}, a=\frac{r_1}{s-n}, s_1=s-n+a, s_2=s-n-a, s_3=a-n-s, s_4=-(s+n+a),
       ~~~~s'=\frac{A}{\sqrt{|\lambda|}} ~\&~
       r'=i\frac{B}{\sqrt{|\lambda|}}.$ The first four entries correspond to $\lambda>0$ and the last two
        correspond to $\lambda<0.$}}
    \end{sidewaystable}}
    \newpage
\begin{center}
{\bf{Appendix \cal{I}}}
\end{center}
 For $B=0$,
$A=\frac{\alpha}{\sqrt{\lambda}}$ the potential of equation
(\ref{eigen2}) and it's energy eigenvalues (\ref{en2}) reduces to
$$V(x)=\left(-\frac{\alpha^2}{\lambda}-\alpha\right) (1+\lambda
x^2)^{-1}+\frac{\alpha^2}{\lambda} \eqno{({\bf{I}}1)}$$
$$
E_n=2n\alpha-n\lambda\label{e41}\eqno({\bf{I}}2)
$$
For $|x| < \frac{1}{\sqrt{\lambda}}$, the potential $({\bf{I}}1)$ can be written as
$$\begin{array}{ll}
\displaystyle V(x)=\left(-\frac{\alpha^2}{\lambda}-\alpha\right)
(1-\lambda x^2
 + \lambda^2 x^4-\lambda^3 x^6+...) +\frac{\alpha^2}{\lambda}\\
 \displaystyle
 ~~~~~~~=\alpha^2 x^2-\lambda(\alpha^2 x^4-\lambda \alpha^2
 x^6+...)+\lambda (\alpha x^2-\lambda \alpha x^4+...)-\alpha
 \end{array}\eqno({\bf{I}}3)
 $$
For $\lambda\rightarrow 0$ the potential reduces to
$$
V(x)=\alpha^2 x^2-\alpha \eqno({\bf{I}}4)
$$
It is clear from $({\bf{I}}4)$ and $({\bf{I}}2)$ that for
$\lambda\rightarrow 0$ the potential (\ref{eigen2}) and the energy
eigenvalues (\ref{en2}) reduces to those of a simple harmonic
oscillator.\\
For $A=\frac{\alpha}{\sqrt{\lambda}}, B=0$ and using the relation
(\ref{e9}) the expression for the wavefunction (\ref{e8}) is as
$$
\psi_n(x)=N'_n (1+\lambda x^2)^{-\frac{2\alpha}{\lambda}}~
\mathcal{H}_n\left(\sqrt{\alpha}
x,\frac{\lambda}{\alpha}\right)\eqno({\bf{I}}5)
$$
where
$$
\begin{array}{ll}\displaystyle
N'_n=\frac{1}{2^n
n!}\left(\frac{\alpha}{\lambda}\right)^{\frac{n}{2}}N_n\\
~~~~=\left[\frac{\alpha^n (\frac{\alpha}{\lambda}-n)
\Gamma(\frac{\alpha}{\lambda}-n+\frac{1}{2})~\Gamma(\frac{\alpha}{\lambda}-n+\frac{1}{2})}{\pi
n!~
2^{2n-\frac{2\alpha}{\lambda}}\lambda^{n-\frac{1}{2}}\Gamma(\frac{2\alpha}{\lambda}-n+1)}\right]^{1/2}
\end{array}\eqno{({\bf{I}}6)}$$
Now for $\lambda\rightarrow 0$ the $\lambda$-deformed Hermite
polynomial becomes the conventional Hermite polynomial $H_n$
{\cite{r18}}. Consequently at $\lambda\rightarrow 0$ limit the
unnormalized wave function given in equation $({\bf{I}}5)$ reduces
to
$$
\psi_n(x) \propto e^{-\frac{\alpha x^2}{2}} H_n(\sqrt{\alpha}x)\eqno({\bf{I}}7)
$$
Using the asymptotic formula  $\Gamma(az+b)\sim
\sqrt{2\pi}e^{-az}(az)^{az+b-\frac{1}{2}}$ (see 6.1.39 of the
ref{\cite{r21}}) in $({\bf{I}}6)$ we have
$$N'_n=\left(\frac{\sqrt{\alpha}-\frac{n\lambda}{\sqrt{\alpha}}}{\sqrt{\pi}2^n n!}\right)^{1/2} \eqno({\bf{I}}8)$$
Therefore from equations ({\bf{I}}7) and ({\bf{I}}8) it follows
that for $\lambda\rightarrow 0$ the wave function given in
equation (\ref{e8}) reduce to that of simple harmonic oscillator.


\newpage
\begin{thebibliography}{99}
\bibitem{r1} L. Dekar et al, J.Math.Phys. {\bf 39} 2551 (1998)\\
L.Dekar et al, Phys.Rev.{\bf A59} 107 (1999)
B. Bagchi et al, Czech.J.Phys. {\bf 54} 1019 (2004)\\
B. Bagchi et al, Mod.Phys.Letts.A {\bf 19} 2765 (2004)\\
J. Yu,S.H. Dong and G.H. Sun, Phys.Letts.A {\bf 322} 290 (2004)\\
J. Yu and S.H. Dong, Phys.Letts.A {\bf 325} 194 (2004)\\
A.Ganguly et al, Phys.Letts.A {\bf 360} 228 (2006)\\
A.D.Alhaidari, Phys.Rev.A {\bf 66} 042116 (2002)\\
S.H. Dong and M. Lozada-Cassou, Phys.Letts.A {\bf 337} 313 (2005)\\
G.Chen and Z. Chen, Phys.Letts.A {\bf 331} 312 (2004)\\
L.Jiang et al, Phys.Letts.A {\bf 345} 279 (2005)\\
C. Gang, Phys.Letts.A {\bf 329} 22 (2004)\\
O. Mustafa and S. Habib Mazharimousavi, Phys.Letts.A {\bf 358} 259
(2006)\\
 S. Cruz y Cruz et al, Phys.Letts.A {\bf 369} 400 (2007)\\
M. Lozada-Cassou et al, Phys.Letts.A {\bf 331} 45 (2004)
\bibitem{r2} R. Koc et al, J.Phys.A {\bf 35} L527 (2002)\\
B. Bagchi et al, Europhys.Letts. {\bf 72} 155 (2005)
\bibitem{r3} B. Roy and P. Roy, J.Phys {\bf A35} 3961 (2002)\\
 R. Koc and M. Koca, J.Phys {\bf A36} 8105 (2003)\\
 C.Quesne, SIGMA {\bf 3} 067 (2007)\\
 S.A. Yahiaoui and M. Bentaiba, preprint, arXiv:0711.2265v1 (2007)\\
S.A. Yahiaoui and M. Bentaiba, preprint, arXiv:0803.4376v1 (2008)
 \bibitem{r4} B. G\"onul et al, Mod.Phys.Letts.A {\bf 17} 2057 (2002)
\\
C. Quesne, Ann.Phys {\bf 321}, 1221 (2006) \\
A. Ganguly and L.M. Nieto, J.Phys.A {\bf A40} 7265 (2007)\\
A.deSouza Dutra et al, Europhys.Letts. {\bf 62} 8 (2003)\\
R. Koc and H. Tutunculer, Ann.Phys.(Leipzig) {\bf 12} 684 (2003)\\
T.Tanaka, J.Phys.A {\bf 39} 219 (2006)\\
V.Milanovic and Z. Ikonic, J. Phys. A: Math. Gen. 32 7001 (1999)
\bibitem{r5} A.R. Plastino et al,
Phys.Rev A{\bf 60} 4398 (1999)
\bibitem{r6} K. Samani and F. Loran, e-print arXive: quant-ph/0302191\\
B.Bagchi et al, J.Phys. {\bf A38} 2929 (2005)\\
S.A.Yahiaoui et al, preprint, arXiv:0704.3425v1 (2007)
\bibitem{r7} B. Roy and P. Roy, preprint, arXiv:quant-ph/0106028\\
A.D. Alhaidary, Phys.Rev.A {\bf 66} 042116 (2002)\\
B. Gonul et al, Mod.Phys.Letts.A {\bf 17} 2453 (2002)\\
O. Mustafa and S. Habib Mazharimousavi, J.Phys.A {\bf 39} 10537 (2006)\\
J.K. Moayedi et al, J.Mol.Struc.THEOCHEM {\bf 663} 15 (2003)\\
K.C. Yung and J.H.Lee, Phys.Rev.A {\bf 50} 104 (1994)
\bibitem{r8} L.Chetouani et al, Phys.Rev.A {\bf 52} 82 (1995)\\
S.H. Dong et al, Int.Jour.Theo.Phys. {\bf 42} 2999 (2003)\\
Y.C. Ou et al, J.Phys.A {\bf 37} 4283 (2004)
\bibitem{r9}  G. Bastard, Wave Mechanics Applied to Semiconductor
Heterostructures (Les Editions de Physique,Les Ulis,France,1988).
\bibitem{r30} M.R. Geller and W. Kohn, Phys. Rev. Lett. {\bf 70} 3103 (1993)
\bibitem{r31} L. Serra and E. Lipparini, Europhys. Lett. {\bf 40} 667 (1997)
\bibitem{r32} M. Barranco, M. Pi, S.M. Gatica, E.S. Hernandez, J.
Navarro, Phys. Rev. {\bf B56} 8997 (1997)
\bibitem{r33} A. Puente, L. Serra, M. Casas, Z.Phys. {\bf D31} 283 (1994)

\bibitem{r34} M.A. Preston, Physics of the nucleus (Addison-Weseley, INC, Reading,
1965), P 210\\
G.H. Wannier, Phys. Rev. {\bf 52} 191 (1937) \\
J.C. Slater, Phys. Rev {\bf 76} 1592 (1949)\\
J. M. Luttinger and W. Kohn, Phys. Rev. {\bf 97} 869 (1955)
\bibitem{r35} G.B.Bachlet, D.M. Ceperley and M.G.B. Chiocchetti ,
Phys. Rev. Lett. {\bf 62} 2088 (1989) \\
W.M.C Foulkes and M. Schluter, Phys. Rev. {\bf B42} 505 (1990)
\bibitem{r14}  C. Quesne and V.M. Tkachuk, J.Phys.A {\bf 37}, (2004) 4267\\
 B. Bagchi, A. Banerjee, C. Quesne and V.M. Tkachuk, J.Phys.A {\bf 38}, (2005) 2929
 \bibitem{r15} A. Mostafazadeh, J.Phys.A {\bf 40} 6557 (2005); Erratum
ibid, {\bf 38} 8185 (2005)
\bibitem{r16} P.M. Mathews and M. Lakshmanan, Quart.Appl.Maths. {\bf 32}
215 (1974)
\bibitem{r29} M. Lakshmanan and S. Rajasekar, "Nonlinear dynamics,
Integrability, Chaos and Patterns", Advanced Texts in Physics 2003
(Springer-Verlag, Berlin)
\bibitem{r17} J.F. Carinena et al, Nonlinearity {\bf17} 1941
(2004)\\
J.F. Carinena et al, Regul.Chaot.Dyn. {\bf 10} 423 (2005)\\
J.F.Carinena et al, J.Phys.Atom.Nucl. {\bf 69}  (2006)
\bibitem{r18} J.F.Carinena et al, Ann.Phys. {\bf 322} 434 (2007)
\bibitem{r19} F.Cooper,A.Khare,U.Sukhatme, Suoersymmetry in Quantum
Mechanics, (World Scientific, 2000)
\bibitem{r20}  G. Levai, Czech.Jour.Phys. {\bf 51} 1 (2001)
\bibitem{r21}  M. Abramowitz and I.A.Stegun, (1964) Handbook of Mathematical
Functions (Dover Publications, INC., New York)
\bibitem{r22} G.Junker, Supersymmetric Methods in Quantum and Statistical Physics
(Springer, 1996)\\
B.K.Bagchi, Supersymmetry in Quantum and Classical Mechanics
(Chapman and Hall/CRC, 2001)
\bibitem{r23}  L.E. Gendenshtein, JETP Letts. {\bf 38} 356 (1983)
L.E. Gendenshtein and I.V.Krive, Sov.Phys.Uspkhi. {\bf 28} 645
(1985)\\
\bibitem{r24}  R. Dutt et al, Phys.Letts.A {\bf 174} 363 (1993)\\
A.Gangopadhyaya et al, Phys.Letts.A {\bf 283} 279 (2001)
 \bibitem{r25} C.M. Bender and S. Boettcher, Phys.Rev.Letts. {\bf 80} 5243 (1998)
%\bibitem{r26}  M. Znojil, Phys.Letts.A {\bf 259} 220 (1999)
\bibitem{r27} B. Bagchi et al, Phys.Letts.A {\bf 269} 79 (2000)

\end{thebibliography}
\end{document}